\documentstyle[prb,aps,twocolumn,amssymb,amsfonts,epsfig]{revtex}

\begin{document}


\title{Ferromagnetic and antiferromagnetic spin fluctuations and superconductivity in the hcp-phase of Fe.}

\author
{T. Jarlborg}

\address{D\'epartement de Physique de la Mati\`ere Condens\'ee,
Universit\'e de Gen\`eve, 24 Quai Ernest Ansermet, CH-1211 Gen\`eve 4,
Switzerland} 

\date{\today}
\maketitle

 High purity iron,
which transforms into the hcp phase under pressure, has recently been reported to be superconducing 
in the pressure range 150-300 kBar \cite{shim}. The electronic structure and the electron-phonon
coupling ($\lambda_{ph}$) are calculated for hcp iron at different volumes. A parameter-free theory
for calculating the coupling constants $\lambda_{sf}$ from ferromagnetic (FM) and antiferromagnetic (AFM) 
spin fluctuations is developed. The calculated $\lambda_{sf}$ are sufficiently large to explain superconductivity
especially from FM fluctuations. The results indicate that superconductivity
mediated by spin fluctuations is more likely than from electron-phonon
interaction.
\begin{abstract}

\end{abstract}


Recent observations of superconductivity in several magnetic or nearly magnetic materials \cite{shim,saxe,pfle} 
have renewed the interest in theories of paramagnon mediated superconductivity \cite{fay,enz,san,mazin:97}. 
Progress in metallurgy has been very important
for these discoveries, since this type of
superconductivity is sensitive to perturbations and impurities. 
In particular, the recent report of superconductivity in the hcp phase of iron, at a 
pressure of about 150 kBar \cite{shim} is interesting,  since it is a non-compound material
where high purity might be easier to achieve. 
A subsequent theoretical work concluded that this superconductivity
could be due to conventional electron-phonon coupling, but that the disappearance of superconductivity
at higher pressure can not be explained by the same means \cite{mazi}. Spin fluctuations, ferromagnetic (FM) or
antiferromagnetic (AFM) have an important role to suppress superconductivity in this scenario.
The alternative explanation, where superconducting pairing is caused by spin fluctuations, has been put forward
for nearly magnetic materials such as ZrZn$_2$ or SrRuO$_3$  \cite{san,mazin:97}. Here we investigate
the possibilities for superconductivity mediated by spin fluctuations in hcp iron.

The electronic structure of bcc and hcp Fe has been calculated using the linear muffin-tin orbital method 
in the local spin density approximation (LSDA) \cite{lmto,lda}. Non-local potential corrections included in the
generalized gradient correction (GGA \cite{gga}) scheme is known to be crucial for a correct description of the fcc vs. bcc
stability of Fe, but other properties are not very different in the two types of density functional 
potentials \cite{moro}. Since we are not focusing on structural stability in this work we use the LSDA potential.
As in ref. \cite{stix} it is found that two AFM configurations are more stable than nonmagnetic or
FM configurations of hcp Fe at high pressure. The first (AFM-I) has opposite polarization on each alternate z-plane layers,
while the second (AFM-II) has opposite polarization on each layer perpendicular to the x-axis. The AFM-I configuration
can be described within the normal hcp unit cell, where the bands are calculated in 252 irreducible k-points.
The AFM-II configuration is described in an orthorhombic unit cell containing 4 atoms, and the bands are calculated
in 200 k-points within 1/8 of the Brillouin zone. The basis set includes states up to $\ell$=3.
The c/a-ratio is fixed to 1.59 for all volumes. This is close to the calculated average value within a wide pressure range \cite{stix}.
The d-band is the dominant character at the Fermi energy ($E_F$) and it is involved in the magnetic orderings.
In the case of spin fluctuations we apply staggered magnetic fields locally on each site. The calculations for the
three types of magnetic configurations are made independently of each other.

The AFM-I configuration develops stable moments for the lattice constant $a >$ 4.81 a.u., and AFM-II for $a >$ 4.61 a.u., 
approximately. The total energy
of the AFM-II state is nearly 2 mRy/atom lower then the AFM-I configuration at the largest volume.
These results are in fair agreement with ref. \cite{stix} despite the use of different potentials. The calculated minimum
in total energy of the bcc structure is found at lattice constant which is $\sim 3$ percent smaller than the experimental one.
The bulk modulus calculated at the experimental lattice constant, about 1.5 Mbar, agree well with experiment, while it
is about 2.2 Mbar at the calculated equilibrium.
Such errors are common when using LSDA for 3d metals. The present results give a lower total energy of the hcp
phase (compared to bcc) for $a < 4.75$.

The electron-phonon coupling parameter $\lambda_{ep}$ is calculated from the electronic structure; 
\begin{equation}
\label{eq:lambda}
\lambda_{ep} = N V^2 / M \omega^2 = N V^2 / K
\end{equation} 
where $N$ is the density of states at $E_F$, $M$ the atomic mass, $\omega$ a phonon frequency and $K$ a force constant.
It is related to the change of total energy $E$ due to an atomic displacement $r$, $ K = d^2E/dr^2$.
The Hopfield parameter $N V^2$ contains
$V$, the Fermi surface average of the matrix element $< \psi_k (dv(r)/dr) \psi_k >$, 
the first order change in potential due to the displacement,
where $\psi_k$ is the wavefunction at $k$.
For elementary metals it can to a good approximation be calculated within the rigid muffin-tin approximation \cite{gg}. 
This matrix element could also be obtained from the change in band energies, $\varepsilon_k$, when the structure
is deformed as in a 'frozen' phonon. The force constant is fitted to the experimental value of the Debye temperature $\Theta_D$
for bcc Fe, and then scaled to hcp and other volumes by the calculated bulk modulus \cite{pict}.

 For FM or AFM spin fluctuations we make a development analogous to the case of
frozen phonon calculations.
Instead of applying a force on an atom to obtain a displacement as in frozen phonon calculations, we 
apply a magnetic field $\xi$ to obtain a magnetic moment $m$. This analogy can be extended to the
calculation of mass enhancement due to spin fluctuations, $\lambda_{sf}$.

\begin{equation}
\label{eq:lam}
\lambda_{sf} = N V^2 / K = N <\psi_k (dv(r)/dm) \psi_k>^2 / (d^2E/dm^2)
\end{equation}

In practice, the calculations are made at two different configurations, one with an applied field
and one without, to give two band structures and two magnetisations. If the changes of the
free energy $(E)$ are harmonic, i.e. quadratic functions of the induced magnetic moment, we have $E_{\xi}=E_0 + J m^2$
where $m$ is the moment (induced by the field $\xi$) per atom.  $E_{\xi}$ is the free energy for the configuration with moment $m$, 
which is the total energy plus the term moment times field.
 The constant $J = \frac{1}{2}d^2E/dm^2 = (E_{\xi} - E_0)/m^2$ is calculated from spin polarized results, where the applied fields range
from 1 to 10 mRy. The local Stoner enhancement, $S$, is
defined as $(\epsilon_{\xi} - \epsilon_0)/\xi$, where $\epsilon_{\xi}$ is the increase of the exchange splitting at an atom
(obtained from the logarithmic derivative
of the d-band) induced by the field $\xi$. The $S$-values are generally smaller for large $\xi$, indicating some aversion against
large moments. Non-harmonic variations of $J$ as function of field are found, especially for FM cases. 
This tendency is less pronounced
for AFM configurations.

The matrix element is estimated from the change in band energies
$ <\psi_k dv \psi_k>^2 = (\varepsilon_k^m-\varepsilon_k^0)^2$, where $\varepsilon_k^m$
is the band energy at point $k$ for the configuration belonging to $m$.  
In the harmonic limit we obtain;
\begin{equation}
\label{eq:lam1}
\lambda_{sf} = N < (\varepsilon_k^m-\varepsilon_k^0)^2 >_{FS} /2(E_m-E_0)
\end{equation}
Here $< ~ >_{FS}$ means a Fermi surface (FS) average.
 
The same formalism can be used for FM and AFM fluctuations. An alternative determination of $\lambda_{sf}$ for the
FM case from the Stoner model and paramagnetic band results \cite{tj86}
gives $\lambda_{sf} = \frac{1}{2} S \bar{S}^2$. The usual Stoner parameters are related by $S=(1-\bar{S})^{-1}$,
and $\bar{S}=NI$, where $I$ is the exchange integral. The two methods give very similar values for $\lambda_{sf}$
as long as the Stoner enhancement is the same in paramagnetic and spin polarized calculations. However,
it is noted that the spin polarized results give somewhat larger Stoner enhancements than the paramagnetic calculations.

Our calculated $\lambda_{ep}$ are of the same order as in ref. \cite{mazi}, but the decrease as function of pressure (P)
is faster. This is mainly caused by our P-dependence of the force constant $K$. The exchange integral I, used for the FM Stoner factor,
calculated from paramagnetic results, varies from 0.065 to 0.067 Ry/atom from the largest to the smallest lattice constant. 
Mazin {\it et. al}
\cite{mazi} obtained 0.075 Ry/atom in their fixed spin method using GGA. This together with a slightly smaller DOS
makes our FM Stoner factors smaller than in ref. \cite{mazi}, but it makes no essential difference for the conclusion
regarding superconductivity based on electron-phonon coupling. In particular, our calculation of $\lambda_{sf}$ 
 without parameters permits to determine the parameter $\alpha$ introduced in ref. \cite{mazi} to fit the transition temperature
near the volume corresponding to $a$=4.76 a.u.. With $\lambda_{sf}=\alpha S$, $\alpha$=0.13 and S=4.6, they obtain $\lambda_{sf}=0.6$,
while we have $\lambda_{sf}=0.54$ at this volume, cf. table 1. The overall fair agreement between the two methods
of estimating $\lambda_{ep}$ and $\lambda_{sf}$ (the latter tends to suppress superconductivity)
leads to similar conclusions for electron-phonon based superconductivity
as in ref. \cite{mazi}. Calculated electron-phonon coupling in nonmagnetic simple cubic Fe is of the same order and 
superconductivity of the order of 10 K has been predicted \cite{free}.

We now turn to the possibility that superconductivity is based on spin fluctuations. A
comparison of the local Stoner enhancements for FM and AFM cases, reveals that the enhancements can be
much larger in the latter cases, in particular for the volumes near AFM instabilities
(cf. Table 1-3). The AFM enhancements decrease rapidly for larger or smaller volumes. 
The enhancements for hcp Fe are calculated within a wide range of volumes, although it should be noted that 
bcc is stable structure at the largest volumes.
The decreasing Stoner enhancements at large volumes,
within the AFM stability region, indicate that it is relatively difficult to increase the local moments further by
applying fields.
 The same effect is probably behind the nonlinear behavior at small volumes, and for FM cases at all volumes, where
there is a saturation of the moments for the largest applied fields. 
 The only exception to this 
behavior at larger fields, is for AFM fluctuations for volumes just smaller than the critical
volume for the AFM instability, where the non-linearity is reversed. The Stoner
enhancements are then larger for the largest field, and J are 'softer'. It is as if a soft AFM mode
 will approach 'zero' field regime near the critical volume for the AFM instability.
The relatively small FM enhancements reflect the fact that hcp Fe is never near a FM instability in this range of
volumes, but a metastable FM state is found at large volume \cite{mazi}.

Despite much larger $S$ for some AFM cases, there are no large differences between the $\lambda_{sf}$ in the
FM and the two AFM cases. One explanation is that the matrix element
for FM fluctuations is more efficient, because as is easy to imagine, a FM field applied in a system
where one band is dominant at $E_F$, will split all bands about equally. The same difference 
$(\varepsilon_k^m-\varepsilon_k^0)$ will appear almost everywhere over the FS. For induced AFM configurations
it is more difficult to visualize these differences. Positive and negative differences are likely, which means
that small differences should exist at some sections of the FS and make the averaged matrix element smaller.

The energy of the fluctuations, $\hbar \omega_{sf}$, is estimated from a Heisenberg model, where the exchange 
integral corresponds to our parameter $J$.
The k-dispersion is $\omega = \omega_{sf} (1-cos (ka))$ for FM, and $\omega = \omega_{sf} sin(ka)$ for AFM fluctuations
  \cite{kittel}.
The characteristic $\omega_{sf}$ to use in an estimate of the superconducting transition temperature
should be some k-point average.
For ferromagnetic fluctuations it has been proposed that $\hbar \omega_{sf} = 1/(4NS)$ \cite{mazin:97}.
This corresponds to $J/4$, since $m^2/(NS)$, the estimated free energy from the Stoner model, is equal to
$J m^2$ in the present approach.

For calculating the superconducting transition temperature we use a weak coupling BCS-like formula
for paramagnon coupling  \cite{bcs,fay}.
\begin{equation}
\label{eq:tcsp}
k_{B}T_{C} = ({\hbar \omega_{sf}}/{1.2}) exp ({-(1+\lambda_{ep}+\lambda_{sf})/ \lambda_{sf}})
\end{equation}
The results using this formula are only approximate, but it turns out that they are of
reasonable amplitude. The results for the three types of fluctuations are shown in the figure. 
The $T_C$ as function of lattice constant in the case of FM fluctuations 
has a triangular shape with a maximum reaching 25 K at the largest volume. However, the stable structure is
bcc for $a$ larger than about 4.75 a.u. according to the calculations. Experimentally, the transition pressure is about 150 kBar,
which would put the transition further away from the bcc equilibrium, somewhere towards 4.6 a.u.. 
From the calculated results one expects a $T_C$ of the order 12 K 
at the hcp transition, while it levels off rather slowly at higher pressure. The difference in P between $a$=4.7 and
4.4 a.u. is of the order 700 kBar, while $T_C$ at the smallest lattice constant still is about 1 K. According
to experiment it is only within a pressure range of 150 kBar that superconductivity is observed.

Superconductivity from AFM fluctuations appears only for volumes (on the nonmagnetic side) near
the AFM instability. The $T_C$'s are lower and the pressure range narrower than in the FM case.
 The AFM-II configuration extends into smaller volumes than AFM-I, further
 away from the bcc-hcp transition. 
The drop in $T_C$ as function of pressure is more significant and corresponds better to experiment than from
FM fluctuations. Whereas the FM fluctuations give a reasonable $T_C$ within a wide range of volumes,
and can develop as soon as the hcp structure is stabilized by pressure,
the AFM fluctuations lead to superconductivity only near the magnetic instability. Thus, if
AFM fluctuations are responsible for the observed $T_C$ it is a coincidence that it
appears just after the bcc-hcp transition.

Two possibilities to explain experiment in terms of spin fluctuations can be discussed. 
First, as superconductivity due to FM fluctuations is sensitive to impurity scattering \cite{foulk}, it is 
possible that lattice defects induced in the pressure experiment near the structural transition
can suppress $T_C$. A decrease of 8-10 K would bring $T_C$ into agreement with the measured one within
a correct pressure range as soon as the hcp phase becomes stable. 
Lattice imperfections like dislocations, interstitial sites, vacancies etc.
may be generated especially near the critical volume for the structural structural transitions, leading
to a gradual onset of $T_C$.
A second possibility is that the concentration of imperfections will suppress $T_C$ from FM fluctuations
completely, leaving the mechanism of AFM-II fluctuations to be responsible for the $T_C$. As was mentioned
the $T_C$ and the pressure range fit experiment reasonable well, but one must understand why
AFM fluctuations can resist better to imperfections than FM ones. Short wave AFM fluctuations
are 'optical' corresponding to $ka=\frac{\pi}{4}$ in the formula for the dispersion. The propagation 
velocity, $\sim d\omega / dk$, is very small for such waves, so that they can be confined between defects.
The scattering with defects is larger for dispersive waves with long wave length. 
The reduction of
$T_C$ compared to $T_C$ itself is proportional to $v_F/(\ell \Delta)$, where $v_F$ is the Fermi velocity,
$\ell$ the impurity scattering length and $\Delta$ the superconducting energy gap \cite{fay}. Our arguments
follow if the velocity of the spin excitations take the place of $v_F$. 
   In both of these possibilities it is expected that
$T_C$ should increase if the concentration of defects can be reduced, since superconductivity based on FM
fluctuations give a larger $T_C$ within a wider range of pressures. Further experiments on high impurity samples
showing superconductivity in a large fraction of the volume should be able to distinguish between the two
possibilities, or if electron-phonon coupling is responsible for superconductivity.





\begin{table}
\caption{Paramagnetic and and FM results for hcp Fe at different lattice
constants. DOS at $E_F$ for the paramagnetic state, Debye temperature, electron-phonon 
coupling ($\lambda_{ep}$), Stoner enhancement, ferro magnetic $\lambda_{sf}$ and exchange parameter $J$. 
}
\begin{tabular}{ccccccc}
 a & $N(E_F)$ & $\Theta_D$ & $\lambda_{ep}$ & $S$  & $\lambda_{sf}$ & J  \\
 a.u. & (Ry atom)$^{-1}$ & K &  &   &  & $mRy/\mu_B^2/atom$   \\
\tableline
4.862 &  21.3 & 460 & 0.37 & 3.2  & 0.76 & 13\\
4.766 &  19.5 & 550 & 0.30 & 2.7  & 0.54 & 16\\
4.670 &  17.8 & 650 & 0.25 & 2.4  & 0.40 & 20\\
4.577 &  16.2 & 750 & 0.21 & 2.2  & 0.32 & 23\\
4.485 &  15.8 & 860 & 0.18 & 2.0  & 0.24 & 29\\
4.396 &  13.6 & 980 & 0.14 & 1.8  & 0.18 & 44\\
\end{tabular}
\end{table}
\begin{table}
\caption{Results of spin-polarized calculations for the AFM-I configuration at different
lattice constants. 
Spontaneous AFM magnetic moment $m$, local Stoner enhancement, exchange
parameter $J = d^2E/dm^2$, and AFM $\lambda_{sf}$.
}
\begin{tabular}{ccccc}
 a & m & S & J & $\lambda_{sf}$  \\
 a.u. & $\mu_B$/atom  & & mRy/$\mu_B^2$/atom     \\
\tableline
4.838 & 1.09 & 9.5 & 25 & 0.11 \\
4.814 & $<$0.02    & 30 & 2.0 & 0.85 \\
4.766 & -    & 13 & 3.5 & 0.40 \\
4.670 & -    & 8  & 8.5 & 0.17 \\
4.577 & -    & 4.8 & 14 & 0.08 \\
4.485 & -    & 3.7 & 19 & 0.05 \\
4.396 & -    & 3.0 & 25 & 0.03 \\
\end{tabular}
\end{table}
\begin{table}
\caption{Results of spin-polarized calculations for the AFM-II configuration at different
lattice constants.
Spontaneous AFM magnetic moment $m$, local Stoner enhancement, exchange
parameter $J = d^2E/dm^2$, and AFM $\lambda_{sf}$.
}
\begin{tabular}{ccccc}
 a & m & S & J & $\lambda_{sf}$  \\
 a.u. & $\mu_B$/atom  & & mRy/$\mu_B^2$/atom     \\
\tableline
4.766 & 0.90 & 6 & 55 & 0.06 \\
4.670 & 0.15 & 10 & 12 & 0.18 \\
4.624 & 0.09 & 15 & 5.5 & 0.35 \\
4.600 & -    & 16 & 4.5 & 0.34 \\
4.577 & -    & 14 & 5 & 0.27 \\
4.485 & -    & 8 & 8.5 & 0.12 \\
4.396 & -    & 5 & 13 & 0.08 \\
\end{tabular}
\end{table}

\begin{figure}[tb!]

\leavevmode\begin{center}\epsfxsize8.6cm\epsfbox{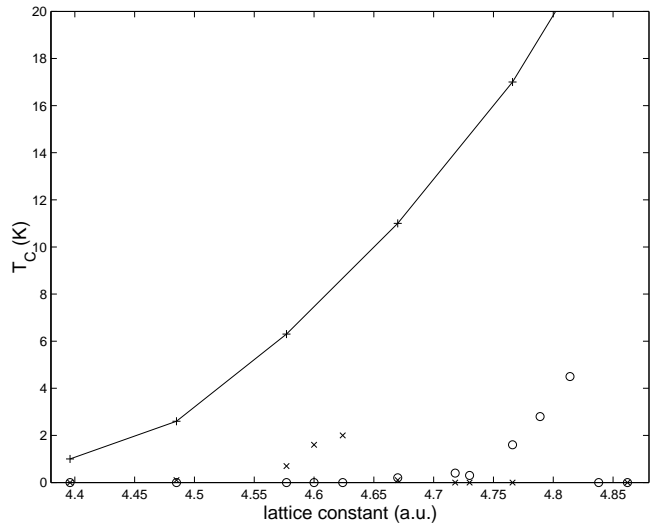}\end{center}
\caption{
 Calculated $T_C$ from FM (full line), AFM-I (circles), and AFM-II (crosses) spin fluctuations in hcp Fe
 as function of the lattice constant $a$. The calculated total energy of the
 bcc structure is lower than for hcp for $a$ larger than about 4.7 a.u..
 The difference in pressure between $a$=4.7 and $a$=4.4 a.u. is about 700 kBar.
Superconductivity  has been observed with $T_C$ below 2 K in a pressure range
of about 150 kBar starting near the bcc-hcp transition pressure \cite{shim}.
 }
\end{figure}

\end{document}